\documentclass[useAMS,usenatbib]{mn2e}

\newcommand{\msun}{M$_{\sun}$}

\newcommand{\msuns}{M$_{\sun}~$}

\newcommand{\mj}{M$_{J}$}
\newcommand{\mjs}{M$_{J}~$}

\newcommand{\mnras}{MNRAS}
\newcommand{\apj}{ApJ}
\newcommand{\apjl}{ApJ}
\newcommand{\aj}{AJ}
\newcommand{\aap}{A\&A}

\newcommand{\apss}{APSS}
\newcommand{\apjs}{ApJS}
\newcommand{\icarus}{Icarus}
\newcommand{\nat}{Nature}
\newcommand{\sovast}{Sov. Ast.}

\newcommand{\lag}{$L_1$}

\newcommand{\nbody}{{\it N}-body~}

\usepackage{graphicx,amssymb,amsmath}

\title[Exoplanets Bouncing Between Binary Stars]
{Exoplanets Bouncing Between Binary Stars}
\author[Moeckel \& Veras]{Nickolas Moeckel$^{1}$\thanks{E-mail:moeckel@ast.cam.ac.uk}, Dimitri Veras$^{1}$\thanks{E-mail:veras@ast.cam.ac.uk}\\
$^{1}$Institute of Astronomy, University of Cambridge, Madingley Road, Cambridge CB3 0HA}

\begin{document}

\date{Accepted XXX. Received XXX; in original form XXX}

\pagerange{\pageref{firstpage}--\pageref{lastpage}} \pubyear{XXXX} 
%\onecolumn

\maketitle

\label{firstpage}

\begin{abstract}
Exoplanetary systems are found not only among single stars, but also binaries of widely varying parameters. Binaries with separations of 100--1000 au are prevalent in the Solar neighborhood; at these separations planet formation around a binary member may largely proceed as if around a single star. During the early dynamical evolution of a planetary system, planet--planet scattering can eject planets from a star's grasp. In a binary, the motion of a planet ejected from one star has effectively entered a restricted three-body system consisting of itself and the two stars, and the equations of motion of the three body problem will apply as long as the ejected planet remains far from the remaining planets. Depending on its energy, escape from the binary as a whole may be impossible or delayed until the three-body approximation breaks down, and further close interactions with its planetary siblings boost its energy when it passes close to its parent star. Until then this planet may be able to transition from the space around one star to the other, and chaotically `bounce' back and forth. In this paper we directly simulate scattering planetary systems that are around one member of a circular binary, and quantify the frequency of bouncing in scattered planets. We find that a great majority (70 to 85 per cent) of ejected planets will pass at least once through the space of it's host's binary companion, and depending on the binary parameters about 45 to 75 per cent will begin bouncing. The time spent bouncing is roughly log-normally distributed with a peak at about $10^4$ years, with only a small percentage bouncing for more than a Myr. This process may perturb and possibly incite instability among existing planets around the companion star. In rare cases, the presence of multiple planets orbiting both stars may cause post-bouncing capture or planetary swapping.
\end{abstract}


\section{Introduction}

Binary star systems which contain exoplanets vary dramatically.
From the first discoveries of exoplanets orbiting stars
in binary systems \citep{butler97} 
to a pulsar planet with a stellar companion \citep{sigurdsson03} 
to the recent circumbinary ``Tatooine''-like planets announced by the 
{\it Kepler} mission team \citep{doyle11s,welsh12s}, exoplanets
have been observed to exist in an intriguing variety of multiple-star systems.
A few years after the discovery of the first multi-planet
system around a main sequence star, $\upsilon$ And \citep{butler99},
\cite{lowrance02} discovered a stellar companion to $\upsilon$ And 
at $\sim 750$ AU.  Subsequently, \cite{curiel11} announced 
a fourth planet orbiting the primary \citep[see also][]{mcarthur10}.  The orbital architecture
of this system may be explained by secular interactions
subsequent to planet--planet scattering and planetary ejection \citep{ford05}, and/or
perturbations from the stellar companion \citep{barnes11}.

Generally, systems with multiple stellar and planetary components
present rich dynamical laboratories which could inform
planetary formation and evolution theories.  Indeed, 
\cite{desidera07} estimate that binary companions within just 
$100-300$ AU are likely to play a role in the formation and evolution of planetary companions. In the Solar neighbourhood, such binaries are prevalent; \citet{raghavan10}, in the most complete survey our local volume within 25 pc, find approximately half of solar-type stars in multiple systems, with a broad peak in the (roughly lognormal) separation distribution stretching from a few to a few thousand au. At separations above a few tens of au, these local multiple systems have the same planet hosting frequency as single stars, setting the stage for potential interactions between stellar and planetary dynamics. An example is \citet{marzari05}, who studied planetary systems undergoing planet--planet scattering from around a star with a binary companion. While the focus of that work was on the orbits of the unejected planets, they found that the binary companion participated directly in the planet scattering, and a few per cent of the ejectees collided with the binary partner; interaction between the binary star and the planets of the most direct sort. 

Here, we investigate planetary transfer between stellar companions
subsequent to planet--planet scattering. In contrast to \citet{marzari05}, we are most interested in the planets that are ejected from their host star rather than the planetary system left behind. Our motivation is best illustrated by considering the motion of test particles in the planar circular restricted three body problem (planar CR3BP). The planar CR3BP, whose early form was established by Euler and Lagrange in 1772 \citep[see][]{barrow-green97}, describes the motion of a massless particle subjected to the force of two massive bodies moving in circular orbits such that all bodies are coplanar.The CR3BP \citep[in its planar form or its extension into three dimensions, the 3D CR3BP, e.g.][]{szebehely67, marchal90} is well suited to describing the motion of an extrasolar planet orbiting one or both stars of a circular binary system; even the most massive planets, at about 13 \mjs \citep{burrows97}, are $< 10$ per cent as massive as the lowest-known exoplanet host star mass (as of 31 January, 2012, $0.13$ \msuns for KOI-691)\footnote{http://exoplanets.org/}.  However, in the vast majority of cases, the planet/star mass ratio is much less than 1 per cent.  Hence, in binary systems with multiple planets, the motion of a planet will be dictated by the CR3BP as long as the perturbations from the other planets are negligible.  This situation arises when one planet is sufficiently far away from the others, which may occur subsequent to planet-planet scattering.

\begin{figure}
 \includegraphics[width=80mm]{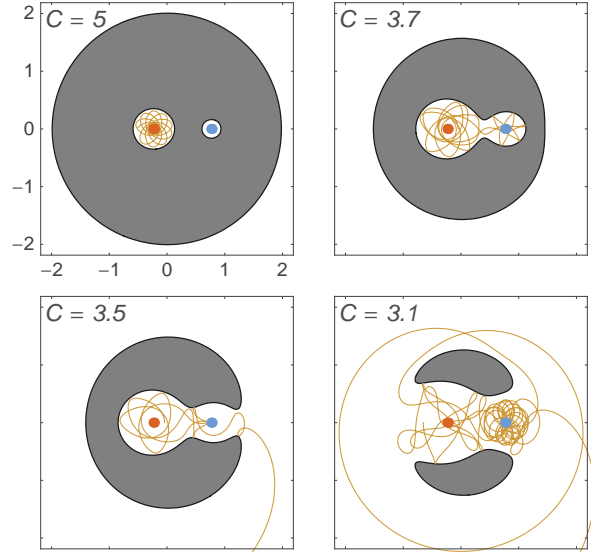}
 \caption{Curves of zero velocity in the dimensionless synodic frame of a circular binary system. In this frame the stars (red and blue dots) are fixed in space. The gray regions show inaccessible regions for test particles with the labeled value of the Jacobian integral. The continuous lines show example orbits at these energies, each originating in the space around the more massive star. At even lower values of the Jacobian constant, the forbidden regions contract around the $L_4$ and $L_5$ Lagrange points before vanishing.}
  \label{jacobianillustration}
\end{figure}

The position and velocity of this planet, as well as the masses and separations of the two stars, determine what regimes of motion are possible for the planet.  \citet{jacobi36} showed that these quantities are related by a constant of the motion, $C$, such that

\begin{equation}
C = x^2 + y^2 + 2 \left(\frac{\mu_p}{r_p} + \frac{\mu_s}{r_s}\right) - \dot{x}^2 - \dot{y}^2 - \dot{z}^2,
\label{JacobianIntegral}
\end{equation}
where the position ($x,y,z$) and velocity ($\dot{x},\dot{y},\dot{z}$) of the planet are computed with respect to the origin of a frame rotating with a mean motion of unity about the center of mass of the stars such that  $\mu_p = G M_p$, $\mu_s = G M_s$, $\mu_p+\mu_s=1$ and where $r_p$ and $r_s$ represent the distance of the planet to the primary and secondary, respectively.  Later, \citet{hill78} linked given values of $C$ with regions of space which are allowed or forbidden.

\begin{figure*}
 \includegraphics[width=180mm]{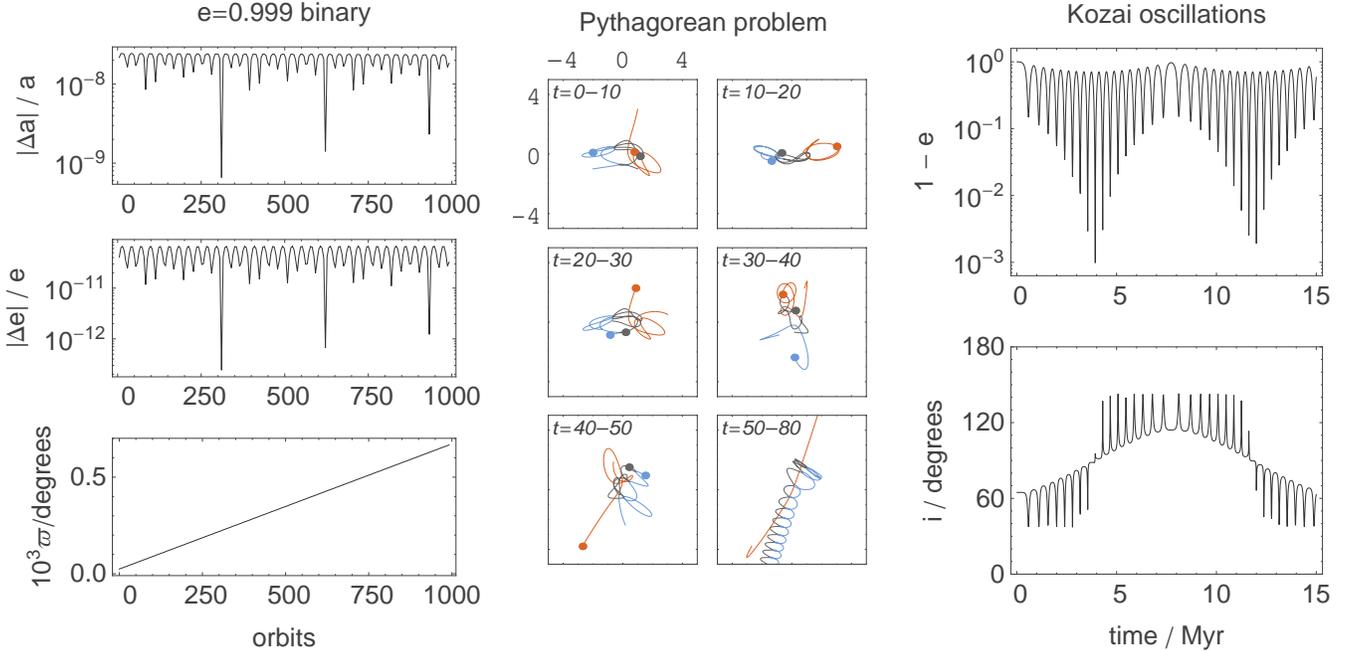}
 \caption{Tests of the integrator used in this paper. {\em Left:} Integration of a planet with eccentricity $e = 0.999$ for 1000 orbital periods. The errors in energy and angular momentum are small and periodic, with no systematic drift. There is, however, a small linear drift in the longitude of perihelion. {\em Middle:} Paths of the stars in the Pythagorean problem through $t=80$. In each panel the positions of the bodies at the end of the time interval are shown with dots. {\em Right:} A test presented in \citet{naoz11}: a system with an inner binary consisting of a 1 \msuns star and a 1 \mjs planet on an $a = 6$ au, $e = 0.001$, $i = 64.7^\circ$ orbit, with a 40 \mjs companion on an $a=100$ au, $e = 0.6$, $i = 0.3^\circ$ orbit. The longitude of pericenter of both companions are initially zero. The inclination and eccentricity of the inner planet are plotted as it undergoes Kozai oscillations, including switches from prograde to retrograde orbits. The agreement to the integrations in \citet{naoz11} is excellent.}
  \label{IntegratorTests}
\end{figure*}

Figure \ref{jacobianillustration} shows examples of these regions in the planar case, i.e. $z=\dot{z}=0$ in equation \ref{JacobianIntegral}.  Suppose a massless planet is initially orbiting the primary.  Then, the trajectory of the planet is given by the continuous line, the primary and the secondary are represented by the large red and blue dots on the left and right respectively, and the gray areas show regions that the planet cannot access.  The relative masses of the stars here are $\mu_s/ \mu_p = 0.3$.  In the $C=5$ plot, the planet is restricted to orbiting the primary.  In the $C=3.7$ plot, the planet may orbit either the primary or secondary, and further, freely travel in between each star's sphere of influence via the connecting region centered on the stars' \lag~Lagrange point.  The bottom two plots illustrate different ways in which the planet can escape the system; the $C = 3.1$ plot demonstrates how the planet can orbit both stars before escaping. While this illustration is in the framework of the planar CR3BP, such transiting orbits exist in the 3D case as well \citep[e.g.][]{gomez04}.

A planet forming around one star of a widely\footnote{`Widely' for these purposes is loosely defined as a system with semi-major axes such that $a_{binary} \gg a_{planets}$, so that the planets are initially orbiting a single star.} separated binary will be in the limit shown in the upper left panel of Figure \ref{jacobianillustration}, although the $C$ value of this planet will constantly change if there are other planets around one or both stars. However, this change is significant only during close approaches with the other planets.  Planet-planet scattering may cause a sudden decrease in $C$, enough to open up the gap around \lag~seen in the $C=3.7$ plot but not enough to enable escape the system through the exterior $L_2$ and $L_3$ Lagrange points, as seen in the $C = 3.5$ and $C=3.1$ plots.  In this case, the planet may `bounce' between the stars.  If the planet is scattered far enough away from the other planets, then $C$ might maintain a value that allows this behavior to continue on appreciable timescales. Here, we explore this possibility, and characterize the prospects and consequences for planetary escape and capture that is internal to a binary stellar system.  

Because planets are not massless, and we wish to treat planet-planet scattering accurately, we cannot rely solely on the analytical structure of the planar or 3D CR3BP for our explorations. In particular, planets often reach ejection velocities via a series of encounters that send them progressively farther from their host star. This rules out introducing test particles with a spectrum of energies in the CR3BP. If ejections were dominantly the result of single encounters, the CR3BP would be more readily and directly applicable, although even then we would need to introduce a stochastic forcing effect within a certain distance of the host star to account for possible interactions with the remaining planets. We therefore turn to direct numerical simulation. Because \nbody codes designed to study planetary systems often rely on the presence of one massive central object that dominates the motion of the other bodies, we constructed a simple and flexible integrator to investigate planetary bouncing. This code and its tests are described in Section \ref{codesection}.  In Section \ref{simulationsection}, we describe ensembles of simulations with the code, and show that bouncing is the dominant behaviour of planets leaving a binary system via planet--planet scattering. We discuss the results and future extensions of this work in Section \ref{discussion}, and then briefly conclude in Section \ref{conclusion}.

\section{The code}
\label{codesection}
The \nbody code we use is built upon a basic fourth-order Hermite integrator, written by P. Hut\footnote{That code is available at http://www.artcompsci.org.}. The straightforward structure of the code makes it easy to adapt for different purposes, although this comes at the sacrifice of fine-tuning that could speed up the code if it were tailored to specific problems. Our modifications and tests are detailed here.

\subsection{Integration Algorthim}
In its implicit form, the Hermite method is time symmetric and exhibits no drift in energy when integrating periodic orbits. In practice, an explicit approximation to the implicit method is usually used, following a PEC (predict--evaluate--correct) cycle. This approach is commonly applied to stellar dynamics problems, where supplementary methods \citep[e.g. KS regularisation,][]{kustaanheimo65} are used to avoid systematic errors in long lived systems with periodic motion. In order for the method to be suitable for studying planetary systems, the implicit version of the Hermite integrator must be used. To achieve this we make two modifications to the base code. First, we iterate over the evaluation of the force and its time derivative and the correction to the predicted values, converting to a P(EC)$^n$ method \citep{kokubo98}. This iteration process converges to the implicit Hermite solution, at the expense of multiple force evaluations. In practice, we use a single iteration, $n=2$.

With fixed timesteps, the P(EC)$^n$ method would yield excellent energy conservation. When timesteps are allowed to be variable, the timesteps must be chosen to be symmetrical functions of the system's state at the beginning and end of the timestep in order to retain the energy conserving features of the Hermite scheme. When all the particles share the same timestep, as in our code, this is straightforward to implement, following the method of \citet{hut95}. This approach requires  iterations over the timestep choice, and we again use one iteration here. A single timestep then involves one iteration for the P(EC)$^n$ integrator, wrapped inside another iteration for the timestep determination. The result is a robust and flexible integrator for small-$N$ systems with no preferred dominant force or geometry. 

The timestep requested by each pair of particles $i,j$ with relative separation ${\bf r_{ij}}$, velocity ${\bf v_{ij}}$, and acceleration ${\bf a_{ij}}$ is proportional to the minimum of their estimated free fall and collision times, the latter under unaccelerated motion: $\delta t_{ij} = \eta {\rm min}[(a_{ij}/r_{ij})^{1/2}, r_{ij}/v_{ij}]$. The system timestep for a given particle state is then the minimum of this quantity over all particle pairs; this estimate is used in the timestep symmetry iteration. In the simulations presented here we set the control parameter $\eta = 0.02$, yielding typical relative energy errors of the order $10^{-8}$, and relative angular momentum errors of the order $10^{-11}$.

\subsection{Tests}
In order to verify the integrator's capabilities to follow planetary dynamics, we have performed several tests. We show the results in Figure \ref{IntegratorTests}. In the first, we integrate a binary system with eccentricity $e=0.999$ for 1000 orbits. We plot the semi-major axis, eccentricity, and longitude of perihelion over the run of the integration. The numerical errors are periodic with minimal systematic drift, with the exception of a small linear error in the longitude of perihelion. Similarly good results are obtained with less extreme eccentricities. The example shown has a secondary to primary mass ratio of 0.1, and similar results are obtained for ratios from unity to approximately zero.

The second test is of the Pythagorean problem \citep{burrau13}. With $G=1$, three bodies of mass 3, 4, and 5 are initially at rest at the corners of a Pythagorean right triangle, each body at the corner opposite the side of the triangle corresponding to its mass. For a detailed account of the complex three body dance that ensues, refer to \citet{szebehely67}, the first authors to resolve the system to its conclusion: after one body is ejected from the system, while the other two recoil as a binary. Those authors used a regularisation scheme to deal with the difficulties of the many close encounters in the system's evolution. Using the parameter for the timestep control that we use throughout this paper, our code integrates the system through virtually the same trajectory as Szebeheley \& Peters find, with relative energy error of the order $10^{-9}$. Following Szebehely \& Peters and \citet{hut95}, we integrate the system with velocities reversed from $t=62$ back to $t=0$. Achieving similar results as those authors, we recover the initial conditions with errors entering at the third decimal place. 

Finally, we integrate a system undergoing Kozai oscillations \citep{kozai62}, leading to flipping of the inclination from prograde to retrograde. The setup is as follows: a 1 \msuns primary is orbited by a 1 \mjs planet with $a = 6$ au, $e = 0.001$, and $i = 64.7^\circ$. There is also a 40 \mjs companion with $a = 100$ au, $e = 0.6$, inclined $65^\circ$ with respect to the inner planet. As the eccentricity and inclination of the companions oscillate, the inner planet oscillates between prograde and retrograde orbits. This is the same system presented in figure 3 of \citet{naoz11}, to which our results can be compared. Those authors integrated the system using a Burlisch-Stoer integrator, and the agreement with that result is excellent.

\begin{figure}
 \includegraphics[width=80mm]{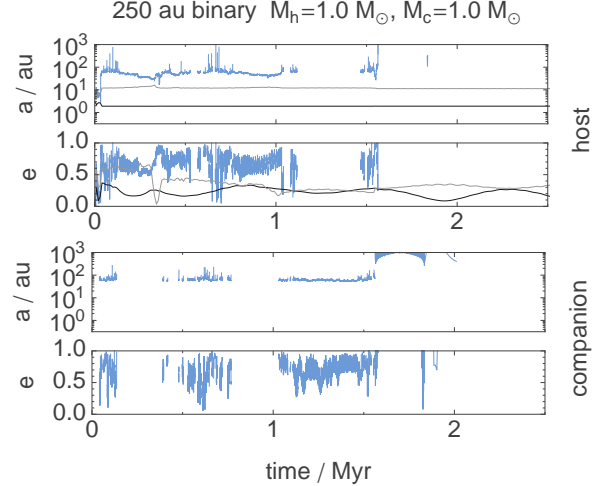}
 \caption{Semi-major axes and eccentricities of planets in a 250 au, equal-mass binary. The orbital elements are plotted relative to whichever star's space the planet is found in at any time, as defined in the text. Top panels are orbits around the host, bottom panels are around the companion.}
  \label{example66}
\end{figure}

\begin{table*}
 \centering
 \begin{minipage}{180mm}
  \caption{Parameters and scattering results for the simulated binary systems. All binaries have zero eccentricity. The majority of scattering planetary systems experience a bouncing planet.
  }
  \begin{tabular}{@{}lcclllllllll@{}}
  \hline
$a$ / au  & $M_h / $\msun & $M_c / $\msun &${\mathcal N}_{scatter}$ & $N_{switch}$ & \multicolumn{3}{c}{ejections} & & \multicolumn{3}{c}{collisions} \\ \cline{6-8} \cline{10-12} \\[-8pt]  
  &                      &                    &     &   &    $N_{total}$ & $n_{switch}$ & $n_{bounce}$ & &  $N_{total}$ & $n_{switch}$ & $n_{bounce}$ \\
    \hline 
 250 & 1.0 &1.0 & 249 &221 &236 & 204 (86$\pm$7\%)& 182 (77$^{+6}_{-5}$\%)& & 84 & 14 (17$^{+6}_{-4}$\%) & 0 (0$^{+2.2}_{-0}$\%) \\
 250 & 1.0 & 0.3 & 231&199 &220 & 192 (87$^{+7}_{-6}$\%)& 138 (63$^{+6}_{-5}$\%)& & 67 & 4 (6.0$^{+4.7}_{-2.9}$\%) & 0 (0$^{+2.7}_{-0}$\%)  \\
  250 & 0.3 & 1.0 & 209 & 200 &226 & 185 (82$\pm$6\%)& 174 (77$\pm$6\%)& & 43 & 10 (23$^{+10}_{-7}$\%) & 3 (7.0$^{+6.8}_{-3.8}$\%)  \\
 \hline
 1000 & 1.0 &1.0& 235 & 185 &215 & 180 (84$^{+7}_{-6}$\%)& 133 (62$^{+6}_{-5}$\%)& & 69 & 1 (1.4$^{+1.4}_{-1.2}$\%) & 0  (0$^{+2.2}_{-0}$\%) \\
 1000& 1.0 & 0.3 & 227& 157 &207 & 156 (75$^{+7}_{-6}$\%)& 73 (35$^{+5}_{-4}$\%)& & 73 & 0 (0$^{+2.5}_{-0}$\%) & 0  (0$^{+2.5}_{-0}$\%) \\
  1000 & 0.3 & 1.0& 165 & 116 &162 & 112 (69$\pm$7\%)& 88 (54$\pm$6\%)& & 24 & 2 (8.3$^{+8.3}_{-5.3}$\%) & 0 (0$^{+7.7}_{-0}$\%)  \\
   \hline     

\end{tabular}
$M_h$ and $M_c$ are the masses of the host and the companion stars.
${\mathcal N}_{scatter}$ is the number of systems that scattered, out of 300 sets of initial conditions. $N_{switch}$ is the number of {\em planets} that cross the \lag~threshold at least once.
In the `ejections' and `collisions' columns, $n_{switch}$ and $n_{bounce}$ are the number that cross \lag~at least once and more than once, respectively; the percentages include Poisson confidence limits corresponding to a $1\sigma$ Gaussian interval.
\end{minipage}
\label{summarytable}
\end{table*}

\begin{figure*}
 \includegraphics[width=180mm]{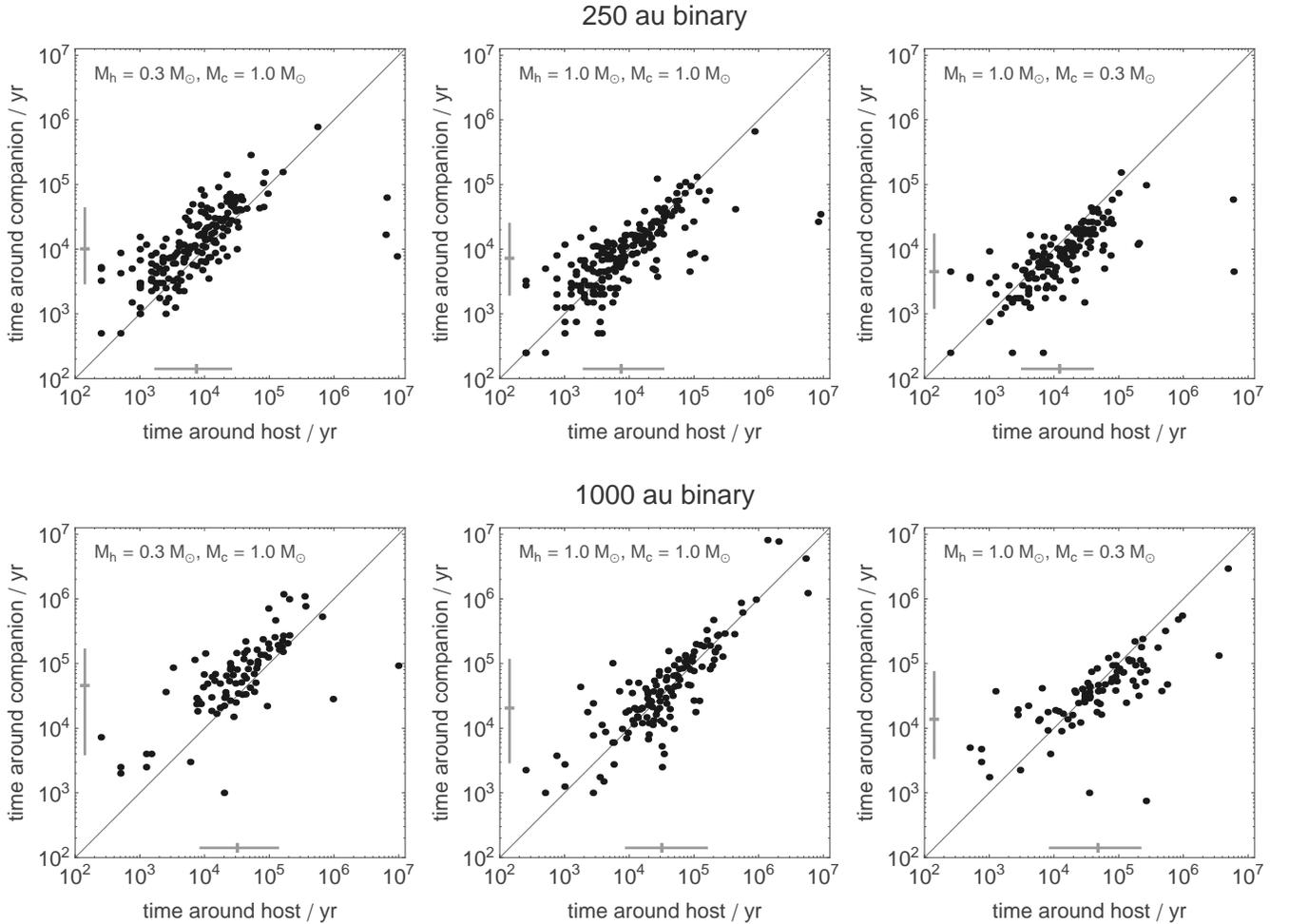}
 \caption{Scatter plots of the time spent in the space around each star for all planets that bounced across the boundary between them. The zero point for the clock is the moment the planet first crosses over the threshold; all the time spent around host prior to being scattered across to the companion is not counted. The slight gridded appearance of points at small times is due to our time resolution for these plots, which is 250 years. The diagonal gray line shows equal time spent around each star. The error bars at the bottom and left show the median of the distribution, and the 16th and 84th centiles.}
  \label{timesafterswitch}
\end{figure*}

\section{The Simulations}
\label{simulationsection}
We performed several sets of 300 simulations each, exploring a limited but illustrative range of binary parameters. Each simulation is run for 10 Myr.

\subsection{Initial Conditions}
We follow a standard planetary setup for scattering simulations, following \citet{marzari02}. Each adjacent pair of planets is separated by $K$ mutual Hill radii, so that the planet $i$ and its outward neighbour $j$ have semi major axes $a_j = a_i + K R_{Hij}$, with
\begin{equation*}
R_{Hij} = \left( \frac{m_i+m_j}{3M_\star}\right)^{1/3} \frac{a_i+a_j}{2},
\end{equation*}
and the chosen value of $K$ determines the instability timescale \citep{marzari02,chatterjee08}. Planetary masses are drawn from a power law mass function with $f(m) \propto m^{-1.1}$, between $0.3$ and $5.0$ \mj.
We take $K=4$ and an innermost semi-major axis of 3 au, which together with our mass spectrum yields an instability timescale of the order $10^4$ yr, although the possibility of long-term stable systems exists with this setup. Eccentricities are initially zero, inclinations are uniformly distributed between $0^\circ$ and $1^\circ$, and all the phase angles are randomized.
Changing the planets' eccentricity and mutual inclinations will alter the instability
timescale, but otherwise yield similar outcomes \citep[e.g.][]{juric08}.

In this initial work, the binary system initially has zero eccentricity, and has semi-major axes of 250 or 1000 au. Small changes to these values are possible as the system evolves and planets are ejected or collide with the stars. The semi-major axis values are chosen to lie within the peak of the local binary separation distribution \citep{raghavan10}, but large enough that the planet formation process around a single star is probably not drastically affected by the presence of its binary companion. The masses of the planet hosting star and its binary partner (henceforth the `host' and `companion', respectively) are taken to be (1.0, 1.0), (1.0, 0.3), and (0.3, 1.0) \msun. We assume coplanarity between the the orbital plane of the binary and the planetary system; this should have only a secondary effect on the results, since planetary ejections from scattering do not occur strictly in the plane of the planetary system.

A planet is considered to be ejected if its distance from the binary center of mass exceeds $5 \times 10^4$ au, a factor of $\sim 2$ less
than a typical distance at which galactic tides can cause escape over a typical main sequence lifetime \citep{tremaine93}
but well beyond the region in which an appreciable number of planets may be scattered onto stable wide orbits \citep{veras09}. Collisions between bodies occur when their radii are detected to overlap; the planets are all 1.3 Jupiter radii, and the stars are 5 Solar radii. Using this large stellar radius enables us to avoid the regime where tidal interactions overly affect the dynamics. Mergers between bodies are momentum and mass conserving.

\subsection{Analysis and Characterisation}

To assign a planet at any instant to one of the two stars, we use the following simple scheme: first, rotate the system about the binary's center of mass so that the stars are on the $x$-axis, and the binary's angular momentum is in the $\hat{z}$ direction. The location of the binary's \lag~Lagrange point is determined, and this is compared to the positions along the $x$-axis of the planets; whichever side of the \lag~divide a planet lies on is its instantaneous host. Planets with distances from the binary center of mass greater than twice the separation of the binary components are tagged so that a planet on a long looping excursion, or one that has been ejected but not yet removed from the calculation, is not counted as being in orbit about either star.

While quite simple, this scheme is motivated by the symmetry of the allowed regions of space in the CR3BP (see Figure \ref{jacobianillustration}). After subtracting off the orbital motion of the binary and appropriately nondimensionalising all velocities, masses, and distances, we also track an approximation to the Jacobian integral of each planet in the 3D CR3BP. In doing this we treat the host and its associated planets as a single body at their barycenter. While this serves to estimate when a planet has sufficient energy to transition between topologies of allowed orbits, it is not exact: the non-zero mass of the planets and small eccentricities in the binary orbit induced when a planet is ejected restrict this to an approximation. With the history of each planet's stellar residence recorded, we tally the number of times each planet crosses the \lag~threshold, as well as the amount of time spent around each star. When a planet crosses the \lag~threshold a single time, we say that it has `switched'; if it crosses the threshold repeatedly, it has begun to `bounce'.

The main questions we wish to answer here are: what fraction of scattered planets transition to the space surrounding the companion star, and how much time do they spend around the companion, or bouncing between the stars?

\subsection{250 au binary results}
We turn first to the results for a 250 au binary. Table \ref{summarytable} provides a summary of the outcomes from our simulations. For the systems with 1 \msuns hosts the number of systems that go unstable and scatter (${\mathcal N}_{scatter}$) during the 10 Myr of our simulations is about 230--250, i.e. 15 to 20 per cent of the runs have not yet scattered\footnote{This fraction is consistent with the results of an identical planetary setup presented in \citet{moeckel12b}, integrated with {\sc mercury} \citep{chambers99}.}. The systems with a 0.3 \msuns host seem to be more stable, with about 30 per cent unscattered. Some difference is expected; the instability timescale is dependent on the mass function \citep{chambers96, marzari02}. Keeping the planetary mass function fixed while reducing the host mass is equivalent to shifting the planets to higher masses. The sense of the change in instability times we see is the same as in previous work: relatively more massive planets are stable for longer times when spaced as fixed multiples of their mutual Hill radii\footnote{
\citet{chambers96} suggest that the instability timescale in a three planet system should vary independently of the planets' mass function when the planets' spacing is scaled as the one-fourth power of their masses, rather than the one-third power in the Hill radius spacing scheme used here.}.

Looking at the number of planets that switch, we see that a large fraction of scattered planets will pass through the companion star's space; the number of switching planets $N_{switch}$ is about 200 for all three binary setups. The overwhelming contribution to this category comes from planets that end up ejected from the binary and eventually removed from the calculation, shown as $n_{switch}$ in the `ejections' columns. For instance, 204 of the 221 planets that switch in the equal-mass binary runs end up ejected, with similar fractions for the other two mass ratios. Crossing the \lag~threshold is symptomatic of a planet that is leaving the system. We illustrate one such system in Figure \ref{example66}, which plots the semi-major axes and eccentricities of the planets with respect to whichever star's airspace they are in at the time. The scattering in this case takes place very early, and the planet shown by the blue line begins to transition back and forth. After an uninterrupted period of about 0.5 Myr around the companion, the final few bounces back to the host bring it into further interaction with the other planets, and its next trip through the companion's space ends with escape from the system. 

Looking further at the ejected planets, note that the majority (about 85 per cent) cross over to the companion on their way out of the system: ejection from the host star occurs predominantly through the space around \lag~rather than the exterior Lagrange point. Furthermore, a majority of ejected planets (about 65 to 75 per cent) not only switch but bounce, traversing between stars multiple times. Escapers are not simply passing through the companion star's space, but rather orbiting there for some time before crossing back to the host, and so onward chaotically until they leave the system or interact further with the rest of the planets.  Planets that ultimately collide with something are less interesting from a switching standpoint; very few collision partners are involved in intrabinary excursions.

The answer to the question of what percentage of ejected planets cross over to orbit the companion is evidently an interestingly large number: something like 85 per cent of escapers will switch. In Figure \ref{timesafterswitch} we address the time they spend in the space around each star by plotting the the time around the host versus the time around the companion for all stars that make the jump and return, with the clock starting from the moment they first pass through the \lag~connection--i.e. we do not count time spent around the host prior to scattering. The diagonal line in the plots shows equal time around each star; the points cluster around this line in a clear correlation. The point lying near 1 Myr around each star in the equal mass case is the system shown in Figure \ref{example66}. A more typical bouncing system is shorter lived, with fewer long term residencies about either star. There is a bias towards spending time around the more massive star in the unequal mass binary cases, but this is not a strong effect. There are a few outliers in each case with large times around the host. These are all cases where a planet made excursions to orbit around the companion, and at some point further interactions with the planets around the host changed the Jacobian integral of the wandering planet, re-isolating it around the host. These adjustments to the Jacobian integral are not possible when the planet is in orbit around the companion.

\begin{figure}
 \includegraphics[width=80mm]{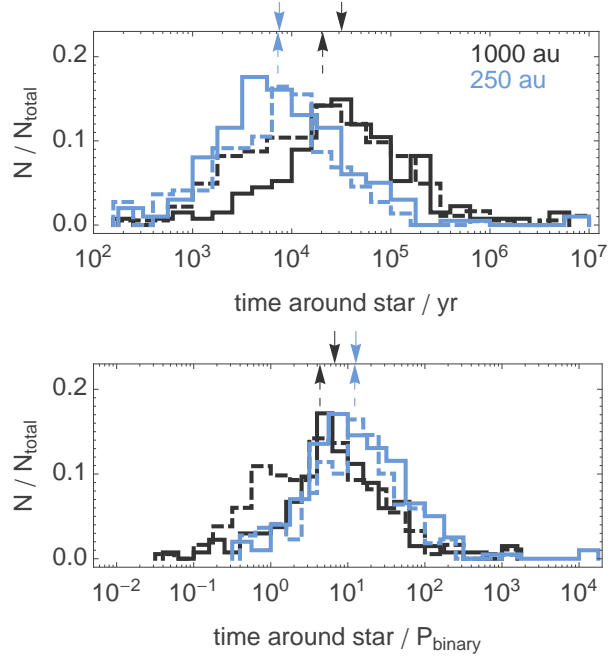}
 \caption{The distribution of times that bouncing planets spend around each star in years (top panel) and binary orbital periods (bottom panel). Time around the host star are shown with solid lines, and time around the companion as dashed lines. The medians of the distributions are shown with arrows at the top of each plot.}
  \label{timerescaled}
\end{figure}

\begin{figure}
 \includegraphics[width=80mm]{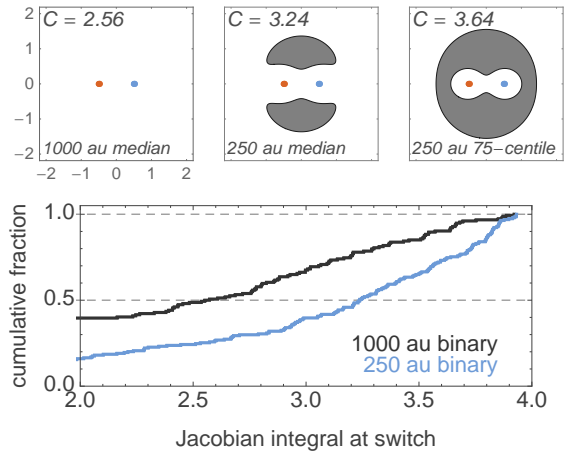}
 \caption{The bottom panel shows the distribution of the Jacobian integral $C$ at the moment a planet first switches to the companion for the equal mass binaries.The top panel shows, in grey, the restricted regions of space in the synodic coordinate system of the CR3BP for three values: the median values of the 1000 and 250 au distributions, and the 75 centile of the 250 au runs.}
 \label{jacobianatswitch}
\end{figure}

\subsection{1000 au binary results}
Overall, the results for the 1000 au binary are quite similar to the 250 au case. This is largely expected; the spacing of the planets, set at a fixed inner radius of 3 au, together with the physical radii of the bodies are the physical quantities that generate the ejection velocities of the planets.
The ratio of the planets' semi-major axes around the host to the binary separation is very small for both of the binary sizes, and the orbits onto which the planets are launched when they begin bouncing should originate from similar points in space when the binary separation is taken as the length scale.The only remaining explanation for changes in the outcomes is the relation between the ejection velocities, a physical quantity set by the planetary dynamics, and the Jacobian integral. A planet with a given {\em physical} velocity and position relative to its host will have a lower Jacobian integral in a more widely spaced binary, since the dimensionless velocity is tied to the binary period.

What differences there are in the outcomes compared to the 250 au binaries are consistent with this picture. The percentage of ejected planets that switch are essentially the same for both binary separations, but the bouncing percentage is lower for the 1000 au binaries. A planet that attains the same ejection velocity from around its host in the 1000 au binary will have a more `wide open' escape through the companion's exterior Lagrange point; this easier escape plausibly reduces the tendency to bounce back to the host. 

The times spent around each star after bouncing begins are longer for the 1000 au systems. If the systems were simple rescalings of an identical dimensionless setup, this difference should be eliminated by rescaling the time to the binary orbital period. We test this with the equal mass binary cases. In Figure \ref{timerescaled} we show the distributions of time spent around each star, which are roughly log-normal, for both binary separations. With time measured in the binary orbital period (the lower panel) the histograms become more similar, although the relative positions of the histogram peaks swap places. Once a planet has been launched by the internal dynamics of the planets around the host into the three-body system consisting of the binary and the ejected planet, the planets bouncing in the 250 au binary are longer lived relative to the binary's dynamical timescale. This is again consistent with the idea that the tighter binary's planets find themselves with a higher Jacobian integral, and less readily able to escape through the region surrounding the exterior Lagrange points.

We can directly test this idea by examining the distribution of the Jacobian integral $C$ at the moment of a planet's first switch. In Figure \ref{jacobianatswitch} we plot this distribution for the equal mass cases for both 1000 au and 250 au binaries. We also show the intersection with the binary's orbital plane of the zero-velocity surface associated with three values of the Jacobian integral: the median value for each binary separation, and the 75th centile of the 250 au binary. This latter value is about the 95th centile of the 1000 au runs. The critical value of the Jacobian integral when the \lag~connection opens is $C_1=4.0$; at $C_{2,3}=3.46$ $L_2$ and $L_3$ open up. As noted earlier, the calculated value of $C$ is not exact, due to small errors introduced by the inconsistencies with the 3D CR3BP in our actual setups. Despite these minor limitations, the difference in the distribution of $C$ is clear; more planets around the 250 au binary have $C_{2,3}<C<C_1$, where access to the outside universe through the exterior Lagrange points is impossible without further perturbations from the remaining planets. In contrast, the great majority of switching planets in the 1000 au binary are free to escape the system.

\begin{figure}
 \includegraphics[width=80mm]{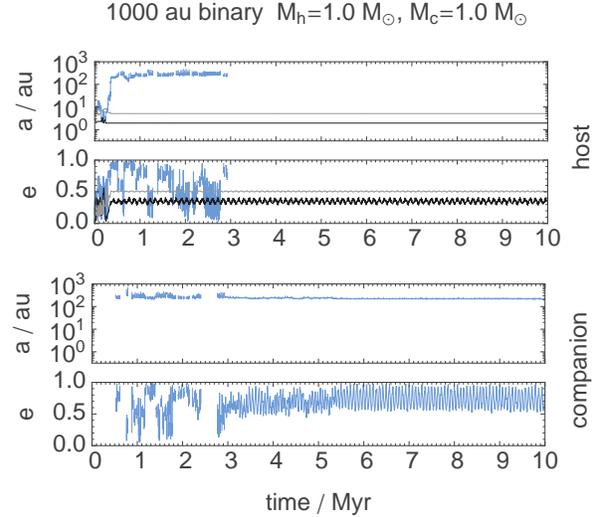}
 \caption{Semi-major axes and eccentricities of planets in a 1000 au, equal-mass binary. The orbital elements are plotted relative to whichever star's space the planet is found in at any time, as defined in the text. The star that is ejected from the host settles into a quasi-stable orbit around the companion. Top panels are orbits around the host, bottom panels are around the companion.}
  \label{example32}
\end{figure}

Since we observe a dimensional universe, the longer physical timescale associated with the 1000 au binaries is of potential interest. In particular there are a handful of cases that spend long periods around the companion, either switching back and forth repeatedly or in rare cases ending up in a quasi-stable orbit about the companion. An example of the latter outcome is shown in Figure \ref{example32}. In this case a planet bounces between stars for several Myr, before settling around the companion at about 3 Myr and staying there for the remainder of the simulation. We stress that this outcome is rare, and is not reliably stable. The semi-major axis of the planet around the companion is about 200 au, and the continued influence of the host star is evident in the eccentricity, which oscillates between about 0.5 and 0.95.

\begin{figure}
 \includegraphics[width=80mm]{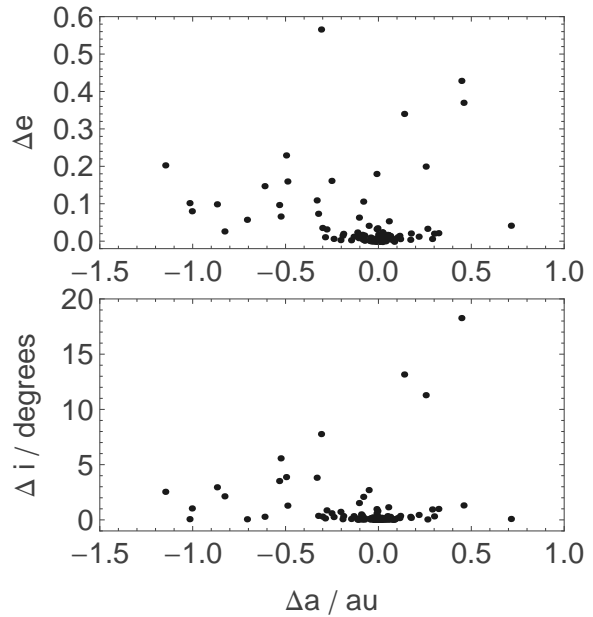}
 \caption{The change in orbital parameters of a single planet around the companion with $e_0=0$, $a_0=10$ au, measured at 10 Myr. These runs were with a 250 au binary and $M_1 = M_2 = 1.0$ \msun.}
  \label{extra1response}
\end{figure}

\begin{figure}
 \includegraphics[width=80mm]{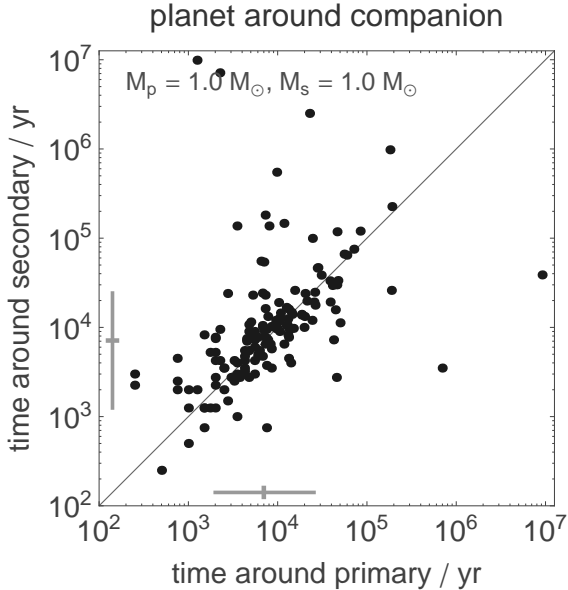}
 \caption{As in Figure \ref{timesafterswitch}, the times spent around each star for bouncing systems in a 250 au binary, but with a single planet orbiting the companion.  Top panels are orbits around the host, bottom panels are around the companion.}
 \label{extra1times}
\end{figure}

\subsection{Response of a planet around the companion}
\label{extraplanetsection}
The orbits of stars passing through the companion's space, even those that are only bouncing for a very short amount of time, are characterised by high eccentricities. Of potential interest is the effect this may have on planets or discs around the companion. We examine this very briefly by conducting simulations identical to the 250 au, equal mass binary runs, but with the addition of a single planet around the companion. This planet has a mass of 1 \mj, a semi-major axis of 10 au, and zero eccentricity. This nearly blank slate of a planetary system records the effect of the wandering planets on planetary systems near the likely terrestrial and giant planet forming region around the companion.

The changes to the semi-major axis, eccentricity, and inclination at 10 Myr of this target planet are plotted in Figure \ref{extra1response}. The majority of systems are relatively unaltered by the intruding planets, but of the order 10 per cent of these planets experience changes in semi-major axis of around 1 au, increases in eccentricity to tens of per cent, and inclination from a few to a few tens of degrees. 

\begin{figure}
 \includegraphics[width=80mm]{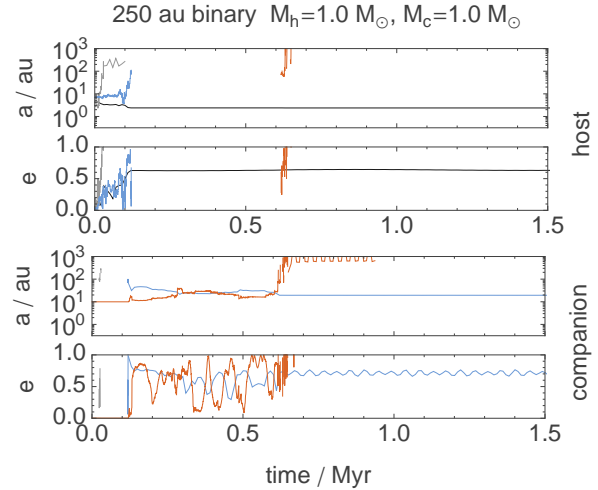}
 \caption{Switching followed by planet displacement in a 250 au, equal-mass binary with a single planet in orbit around the companion. The orbital elements are plotted relative to whichever star's space the planet is found in at any time, as defined in the text. The star that is ejected from the host interacts with a planet originally around the companion, before ejecting it and settling into a stable orbit around the companion. The oscillations in that planet's eccentricity are due to interactions with the host star.}
  \label{example180}
\end{figure}

These changes to the single planet's orbit imply some change to the orbit of the intruding planet, possibly altering the time spent around each star. In Figure \ref{extra1times} we show these times. There is very little change, except for the presence of a few systems that spend more time around the companion, filling in the space above the diagonal reference line. These are not very notable, with two exceptions. Twice the planet ejected from the host displaces the planet around the companion, taking its place and ejecting the companion's original planet from the system. These are the only two cases where the planet originally around the companion is ejected. One of these runs is shown in Figure \ref{example180}. The interactions between the incoming planet and the companion's planet occur over about 0.5 Myr, culminating in the escape of that planet after briefly crossing over to the host system.

\subsection{Non-zero binary eccentricity}
\label{eccentricitysection}
For simplicity, we have dealt exclusively with zero eccentricity binaries in this first work. The eccentricity distribution of long-period binaries spans all values; \citet{raghavan10} find a flat distribution up to $e \sim 0.6$, with the drop off above that value probably due at least in part to observational bias. Eccentricity in the binary may expected to change the results of this work; Hill stability of a test particle is a function of the massive bodies' eccentricity \citep{marchal82,holman99,szenkovits08}, and this dependence could alter the timescale or qualitative nature of the scattered planet's trajectory in the binary.
While will we defer a more complete study of the effect of eccentric binaries on these results to other work, the results of one example case hint at the magnitude and character of the differences to expect.

We re-ran the 250 au equal mass case with a binary eccentricity of 0.75. The first point to note is that the closer approaches of the stars perturbed the planetary systems so that only 14 cases remained unscattered after the 10 Myr runs. This destabilizing effect is described in \citet{holman99} and noted in \citet{marzari05}. Similarly to the latter authors, we find more planets ejected (per scattered system) at higher eccentricity; $\sim 1.25$ per system compared to $\sim 0.95$ in the zero eccentricity case. The fraction of ejected planets that cross L1 is similar, 89 per cent with $e=0.75$. The number of bouncing planets is virtually the same at 76 per cent compared to 77 per cent in the circular runs. The times spent around the host and companion are smaller, by about a factor of 2. The introduction of moderate eccentricity appears to preserve the qualitative and many quantitative aspects of this work, though the detailed dependence of these results on the binary eccentricity deserves further attention. We note that as the binary periastron decreases to be comparable to the planet's semi-major axis the companion may take a more direct role in the scattering \citep{marzari05}, perhaps leading to a qualitative shift in the outcomes.

\section{Discussion and Future Directions}
\label{discussion}
A natural extension to this work would be to remove various restrictions of the CR3BP. For example, if the two stars are not bound to one another, but are hyperbolically scattering off of each other, then one may utilize the parabolic or hyperbolic restricted three-body problem \citep{lukyanov10,sari10} in order to determine the resulting motion of the planet.  However, as touched upon in section \ref{eccentricitysection}, a closer extension to our study would involve removing the circular orbit restriction but keeping the stars bound to each other. If the stellar masses orbit one another on an unchanging, eccentric orbit, then the Jacobian integral no longer exists but the 5 Lagrangian points still do \citep{szebehely64,szebehely67c}.  The coordinate system traditionally adopted for this situation is nonuniformly rotating and pulsating; the zero-velocity surfaces change with time. The consequence is a system which is difficult to describe in simple closed analytical forms \citep[e.g.][]{contopoulos67,delva79,gawlik09,hou11} but can still admit escape and capture \citep[e.g.][]{bailey72,huang83,llibre90,mako04}.

In all of these situations, a star cannot capture a planet and retain it over the star's main sequence life without some additional dissipative force. In the traditional photogravitational restricted three-body problem \citep{radzievskii50}, this force arises from the star's radiation on a massless dust particle. For a planet in a binary system, however, a non-negligible force might arise from additional planets (as briefly discussed in Section \ref{extraplanetsection}) or from mass loss or gain from the stars or planet \citep{bekov88, bekov91, lukyanov09,letelier11}. This situation is particularly relevant {\it after} one or both of the stars have left the main sequence, however the bouncing timescales we find here (typically less than a few Myr) are shorter than stellar evolution timescales. For a discussion of planetary transfer in post main sequence binary systems, see Kratter \& Perets (2012, in preparation).

A more likely dissipative mechanism in the planet formation context is interactions with discs, either of protoplanets, debris, or gas. Scattering can occur when significant amounts of gas are still around the star \citep{moeckel12b}, and if the two stars of the binary are at similar stages in their lives a planet scattered over to the companion star could encounter a significantly dissipative medium. If dissipation is sufficient to strand the marauding planet around the companion star, the orbit would likely be a wide one, offering a way to produce giant planets in the inner regions of a disc and send them off to a wider orbit around a binary companion. This sort of scenario is probably most efficiently explored via a dissipative prescription of some sort in the CR3BP rather than direct integration of a hydrodynamic disc. Finally, while we have focused on planets themselves scattering across to the companion star in this work, other scattering events are likely in planet formation. Scattered planetesimals and exoplanetary analogues to our own outer Solar System structures (e.g. the Kuiper belt, or scattered disc objects) may be able to participate in jumping between stars in a similar fashion to what we have shown here.

The results of this work are arguably entertaining, but are they observable? The short timescales of active bouncing that we observe would suggest not; the consequences of planets bouncing back and forth are likely to be seen in the extensions discussed above, particularly a dissipative environment around the companion, which could lead to permanent capture of the scattered planet, or the introduction of a planetary system around both stars.
The cursory examination of the response of a planet around the companion that we performed here shows that a small but non-negligible percentage of compact planetary systems could be perturbed by the intrusion of a scattered planet. There are many possible consequences of this: changes to the eccentricity of outer planets can propagate inwards through the system\citep{zakamska04}; two-planet systems in formally Hill stable configurations could be pushed into unstable architectures; mean motion resonances could be broken; and richer systems than the single planet that we simulated could be induced into scattering themselves, resulting in wholesale transfer of planets between the two stars. These are the topics of ongoing investigations.

\section{Conclusions}
\label{conclusion}
Via direct integration of a planetary system around one member of a circular binary system, we have examined the idea that planets can be sent bouncing between the stars during planet--planet scattering. The internal dynamics of the planetary system can increase the energy of one or more planets, enabling it to cross over to the companion star from its host. We find that this is in fact a dominant outcome of scattering in a binary: about 70 to 85 percent of planets ejected from binaries in the separation range 250 to 1000 au will spend time around both stars before leaving the system entirely, and about 45 to 75 percent will bounce repeatedly back and forth. The amount of time spent bouncing is astronomically short: a few to tens of binary periods, although with some systems surviving hundreds to thousands of periods.
We propose that the impact of this bouncing will most likely be seen in the response of the companion star's own planetary system, or with the inclusion of un-modeled dissipative forces, such as discs, around the companion.

\section*{Acknowledgments}
Our thanks to Dan Fabrycky for a clear and useful referee's report, and to Kaitlin Kratter and Cathie Clarke for comments on the draft.
Large portions of this work were performed using the Darwin Supercomputer of the University of Cambridge High Performance Computing Service (http://www.hpc.cam.ac.uk/), provided by Dell Inc. using Strategic Research Infrastructure Funding from the Higher Education Funding Council for England.

\bsp

\label{lastpage}

\end{document}